# Finite element solution of multi-scale transport problems using the least squares based bubble function enrichment

A. Yazdani [a], V. Nassehi [b1]

[a] Cranfield University, School of Applied Sciences, Cranfield, MK43 0AL, UK
[b] Loughborough University, Department of Chemical Engineering, Loughborough, LE11 3TU, UK

**Abstract**

This paper presents an optimum technique based on the least squares method for the derivation of the bubble functions to enrich the standard linear finite elements employed in the formulation of Galerkin weighted-residual statements. The element-level linear shape functions are enhanced with supplementary polynomial bubble functions with undetermined coefficients. The best least squares minimization of the residual functional obtained from the insertion of these trial functions into model equations results in an algebraic system of equations whose solution provides the unknown coefficients in terms of element-level nodal values. The normal finite element procedures for the construction of stiffness matrices may then be followed with no extra degree of freedom incurred as a result of such enrichment. The performance of the proposed method has been tested on a number of benchmark linear transport equations with the results compared against the exact and standard linear element solutions. It has been observed that low order bubble enriched elements produce more accurate approximations than the standard linear elements with no extra computational cost despite employing relatively crude mesh. However, for the solution of strongly convection or reaction dominated problems significantly higher order enrichments as well as extra mesh refinements will be required.

## 1. Introduction

A major problem associated with the finite element solution of the differential equations describing different types of transport phenomena (e.g. convection, diffusion and reaction) is that such solutions, in general, fail to capture non-smooth behaviour of the unknown variables. In particular, if the field variables change rapidly within the thin internal or boundary layers, standard numerical schemes lead to inaccurate and unstable results observed as spurious oscillations. For example, standard Galerkin finite element schemes are known to yield oscillatory solutions for convection or reaction-dominated multi-scale transport problems [1-3]. Multi-scale phenomena are those in which the field variables show different orders of magnitude in the scale of their variations. Fine scale variations usually demonstrate their affects during the abrupt changes in the behaviour of the field unknowns within a narrow layer in the problem domain. In such situations, traditional discretization methods produce relatively accurate results only if excessively fine computational mesh is used. In the context of finite element schemes to satisfy this requirement the size of the elements should be smaller than the thickness of layers in which abrupt changes (i.e. fine scale variations) occur. This often results in a large system of algebraic equations whose solution may either become inaccurate or even worthless due to the computational error build-up, or it may involve high computational cost to say the least.

Development of variational multi-scale methods [4-6] enabled researchers to cope with such problems beyond the power of classical finite elements. In particular, these techniques can be used to solve finite element problems in which the chosen discretization level does not satisfy the stability conditions. The variational multi-scale method provides a robust basis to overcome the problems associated with the use of Galerkin method that employs standard linear shape functions in the presence of multi-scale phenomena. In a combined two-scale approximation, the element-level variations of an unknown ($u$) is expressed in terms of two components as $u = u_l + u_b$ where $u_l$ is an approximation based on the standard piecewise linear functions and $u_b$ is the function which locally solves the residual differential equation subject to homogeneous boundary conditions. This residual equation is generated from the insertion of $u - u_l$ into the original model equation. Component $u_b$ is called "Residual-Free Bubble" (RFB) which satisfies the residual equation strongly. Discretization of the problem domain and applying the described process at an elemental level provides a set of enriched shape functions which can be used to construct a robust finite element scheme. An important property of the bubble functions is that they are defined within each element and vanish at the element boundaries [7-10]. Justification for such a definition is that the irresolvable fine-scale behaviour exists locally within each element but not at the element boundaries. The RFB method, offers a powerful theoretical approach for the solution of multi-scale problems in which both fine and coarse scale variations are taken into account [4-8]. Another advantage of this method is that any additional degree-of-freedom associated with RFB implementation can be eliminated from the global set of equations by using a technique known as the static condensation [11]. Therefore extra accuracy is achieved without using elements with more nodes than linear elements. Typically, the RFB process starts with an analytical solution of the residual equations subject to homogeneous boundary conditions. Achieving this, however, may be difficult in the problems described by complex partial differential equations. Another issue is that the analytically derived solutions may be represented as complicated mathematical expressions (e.g. trigonometric or hyperbolic series) which cannot be directly used in the evaluation of numerical integrals normally appearing in a finite element solution procedure. Therefore, in order to overcome such loss of flexibility it is more practical to convert the residual-free bubble functions to simpler forms by using Taylor series approximations instead of original analytical solutions or/and semi-discrete methods [3, 12-13].

---

[1] Corresponding author: V.Nassehi@lboro.ac.uk (V. Nassehi).

In this paper, a least squares method has been used to derive the bubble functions required for the enrichment of standard linear finite elements in Galerkin weighted-residual statements. Such bubble functions provide the best polynomial fit for the element-level residual-free bubble function in as a uniform approximation (in contrast to the described point-wise approximations [3,12-13]). The standard linear shape functions are enriched with a high order supplementary polynomial bubble function with unknown coefficients to form a trial solution. Insertion of the enriched approximating function into the operator equation results in a residual term. Following this step a residual functional is constructed using the normal weighted residual technique. Such a residual functional will always be convex and hence its minimum can be found by taking its partial derivates with respect to the unknown coefficients and setting them equal to zero. This results in an algebraic system of equations. The solution of this system yields the unknown coefficients of enriched functions in terms of element-level nodal values. The finite element procedures to construct the local stiffness equations and assembling the global stiffness matrices will follow normally. One of the advantages of the described procedure over the RFB method is that the laborious (or in many cases impossible) manipulations needed to derive an analytical solution for the local equations are avoided. Therefore, the proposed scheme is, potentially, of greater practical value as it can be completely automated in a finite element programme. In what follows, the construction of the least squares bubble enhanced elements has been demonstrated and the performance of the proposed scheme has been tested on a number of benchmark linear transport equations. The results obtained using this scheme have been compared against the exact solution and numerical results yielded by a standard scheme.

## 2. Residual free bubble functions

The RFB method is based on the analytical solution of the model differential equation within each element using homogeneous boundary conditions [14]. Consider the following boundary value problem

$$\begin{cases} Lu = f \ in \ \Omega \\ u = 0 \ on \ \partial\Omega \end{cases} \tag{1}$$

where $\Omega$ is a domain in $R^n$ with a boundary of $\partial\Omega$, $L$ is a linear differential operator, $u$ is an unknown (scalar or vector valued) function and $f$ represents a source or a sink. We assume that $L$ is such that the problem is well posed i.e. it has a unique solution continuously dependent on the initial values. Assume a partition of $\Omega$ into a set of admissible elements $\Omega_e$ such that no two elements overlap and the union of all such elements is $\Omega$. The approximation space $V_h$ is chosen from a finite dimensional space depending on the partition that satisfies $V_h \subset V$, where $V_h$ is the space of the functions in which a solution to the continuous variational problem is sought. Here, the partition diameter is $h = max\{diam(\Omega_e)\}$. The Galerkin method seeks $u \in V_h$ such that

$$\langle Lu|v \rangle = \langle f|v \rangle \quad \forall v \in V_h \tag{2}$$

where $\langle .|. \rangle$ is a bilinear form of the variational problem given as Equation (1). To enrich the elements used in the standard Galerkin scheme with bubble functions, each $u \in V_h$ is taken as the sum of standard piecewise linear parts and bubble functions as $u = u_l + u_b \ \forall u \in V_h$ where $u_l$ and $u_b$ are the linear and the bubble components, respectively. The bubble function is selected to be such that it vanishes on the element boundaries. Fin the residual free bubble function method, $u_b$ needs to strongly satisfy the residual equation within each element, therefore

$$\begin{cases} Lu_b = -Lu_l + f \text{ in } \Omega_e \\ u_b = 0 \quad \text{on } \partial\Omega \end{cases} \quad (3)$$

As the bubble function vanishes on each element boundary static condensation [11,15] can be used which leads to the following selection of $v = v_b|_{\Omega_e}$ in Equation (2). Hence, the weak representation of Equation (3) is

$$\langle Lu_l + Lu_b | v_l \rangle + \langle Lu_l + Lu_b | v_b \rangle = \langle f | v_l \rangle + \langle f | v_b \rangle \quad for \quad v_l + v_b \in V_h \text{ on } \Omega_e \quad (4)$$

Therefore it is sufficient to find $u = u_l + u_b$ in such a way that

$$\langle Lu_l + Lu_b | v_l \rangle = \langle f | v_l \rangle \quad \forall v_l \in V_h \text{ on } \Omega_e \quad (5)$$

The approximate solution of Equation (3) can be expressed as

$$u = u_l + u_b = \sum_{i=1}^{n} u_i(\phi_i + \psi_i) + \phi_f \quad (6)$$

where $\phi + \psi$ are the linear and bubble functions, respectively and $n$ is the number of nodes per element. Evaluation of the unknown nodal values $u_i$ is achieved by a routine finite element solution process.

## 3. Least squares bubble function for one-dimensional transport model

From a practical point of view, polynomials, in general, present the preferred choice for approximating functions due to the level of flexibility they provide towards mathematical operations such as differentiation, integration and arithmetic calculations. Therefore polynomials generated by Taylor series expansion of polynomial bubble functions instead of their original forms have been widely used to construct finite element schemes for multiscale problems [12-13]. While Taylor series usually provide good point-wise approximations for the estimated functions defined over points within a small radius of convergence, they usually fail to capture the function variations (expressed as derivatives) at points further away from the centre of the neighbourhood of convergence. In contrast, polynomials generated by the least squares method provide more uniform approximations (that converges in the mean) on a fixed interval or within the rectangular finite element grids. The technique presented in this section uses this property of the least squares method to derive polynomial bubble functions. An important property of this approach is that once the unknown polynomial coefficients are calculated, they can be used directly in the computer programme routines to calculate the element-level nodal values.

The general steady state model for transport processes incorporating convection, diffusion and reaction phenomena is expressed by equation $\varepsilon \Delta u + \langle \vec{\kappa} | \nabla u \rangle + \lambda u = f \text{ in } \Omega$ subject to the prescribed boundary conditions. The simplified one-dimensional version of this equation is expressed by

$$\begin{cases} \varepsilon u'' + \kappa u' + \lambda u = f \text{ in } \Omega \\ u(a) = \alpha, u(b) = \beta \text{ on } \partial\Omega \end{cases} \quad (7)$$

in which the assumption of $f = 0$ (no sinks or sources) has been made for the sake of simplicity. The problem domain can be discretized into $N$ sub-intervals of the length $l_j = x_{j+1} - x_j$ where $0 = x_0 < x_1 < \cdots < x_j < x_N = 1$ for $j = 1, \ldots, N$. Consider the local approximation for $u$ in the $j$-th subinterval $I_j = [x_j, x_{j+1}]$ is given by

$$\tilde{u}(x) = \frac{x - x_j}{l_j} u_{j+1} + \frac{x_{j+1} - x}{l_j} u_j + B_{2,j}(x) \quad (8)$$

where $B_{2,j}(x) = c_j(x_{j+1} - x)(x - x_j)$ is the supplementary quadratic bubble function over the element $I_j$ and $u_j = u(x_j)$ and $u_{j+1} = u(x_{j+1})$ are the nodal values. The bubble function $B_{2,j}(x)$ is defined to be zero on the boundaries and outside of $I_j$. The bubble coefficient $c_j$ shall be determined in such a way that provides the most accurate approximation at the element level, without the requirement for the introduction of additional midpoints. With no major loss of generality, it can be assumed that the calculations are taking place over a master element represented by a standard interval $[0, l]$. Rewriting Equation (8) provides

$$\tilde{u}_j(x) = \frac{l-x}{l}u(0) + \frac{x}{l}u(l) + c_j x(l-x) \tag{9}$$

The residual term $R_B$ within the local interval $I_j$ results from the insertion of Equation (9) into Equation (7), as

$$R_B = -2\varepsilon c_j + \kappa\left(\frac{u_l - u_0}{l} + c_j(l-2x)\right) + \lambda\left(\left(\frac{l-x}{l}u_0\right) + \frac{x}{l}u_l + c_j x(l-x)\right). \tag{10}$$

We define the residual functional, essential in finding the unknown coefficient $c_j$, as given below [16]

$$J_B = \int_0^l R_B^2\, dx \tag{11}$$

The functional $J_B$ is a convex function where its minima can be found by differentiating with respect to $c_j$ and setting each partial derivative to zero

$$0 = \frac{\partial J_B}{\partial c_j} = \int_0^l 2R_B \frac{\partial R_B}{\partial c_j}\, dx. \tag{12}$$

Expression (12) may be rearranged to find unknown coefficient $c_j$ (in general, a system of algebraic equations generated depending on the number of unknown coefficients in the polynomial bubble functions). The computed value of $c_j$ following the rearrangement of (12) is

$$c_j = \frac{5}{2}\left(\frac{(-\lambda^2 l^3 + 12\varepsilon\lambda l)(u_l + u_0) + 24\varepsilon\kappa(u_l - u_0)}{\lambda^2 l^5 - 20\varepsilon\lambda l^3 + 10\kappa^2 l^3 + 120\varepsilon^2 l}\right). \tag{13}$$

By replacing $c_j$ into the Equation (9), one obtains

$$\begin{cases} \tilde{u}_j(x) = \frac{l-x}{l} + (A-B)x(l-x)u(0) + \frac{x}{l} + (A+B)x(l-x)u(l) \\ A = \frac{5}{2}\left(\frac{(-\lambda^2 l^3 + 12\varepsilon\lambda l)}{\lambda^2 l^5 - 20\varepsilon\lambda l^3 + 10\kappa^2 l^3 + 120\varepsilon^2 l}\right) \\ B = \frac{5}{2}\left(\frac{24\varepsilon\kappa}{\lambda^2 l^5 - 20\varepsilon\lambda l^3 + 10\kappa^2 l^3 + 120\varepsilon^2 l}\right) \end{cases} \tag{14}$$

The weighted residual statement of Equation (7) corresponding to the weight function $w(x)$ can be written as

$$0 = \int_0^l w(x)(\varepsilon\tilde{u}_j''(x) + \kappa\tilde{u}_j'(x) + \lambda\tilde{u}_j(x))\, dx. \tag{15}$$

Consequently, integration by parts can be applied

$$-\varepsilon\int_0^l w'(x)\tilde{u}_j'(x)dx + \kappa\int_0^l w(x)\tilde{u}_j'(x)dx + \lambda\int_0^l w(x)\tilde{u}_j(x)dx = -\varepsilon\{w(x)\tilde{u}_j'(x)\}_0^l \tag{16}$$

The Galerkin weight functions are selected to be

$$\begin{cases} w_0(x) = N_I(x) = \frac{l-x}{l} + (A-B)x(l-x) \\ w_0(x) = N_{II}(x) = \frac{x}{l} + (A+B)x(l-x) \end{cases}$$

for which the associated stiffness matrix of the weighted residual statement is

$$\begin{bmatrix} -\varepsilon \int_0^l N_I'N_I'dx + \kappa \int_0^l N_I N_I'dx + \lambda \int_0^l N_I N_I dx & -\varepsilon \int_0^l N_{II}'N_I'dx + \kappa \int_0^l N_{II} N_I'dx + \lambda \int_0^l N_{II} N_I dx \\ -\varepsilon \int_0^l N_{II}'N_I'dx + \kappa \int_0^l N_{II} N_I'dx + \lambda \int_0^l N_{II} N_I dx & -\varepsilon \int_0^l N_{II}'N_{II}'dx + \kappa \int_0^l N_{II} N_{II}'dx + \lambda \int_0^l N_{II} N_I dx \end{bmatrix} \begin{bmatrix} u_0 \\ u_l \end{bmatrix} = \begin{bmatrix} -\varepsilon \{N_I \phi\}_0^l \\ -\varepsilon \{N_{II} \phi\}_0^l \end{bmatrix} \quad (17)$$

where $\phi$ is the boundary line term. Substitution of shape functions into the above matrix system of (17) and evaluation of the integrals and boundary line terms provide the following element-level stiffness matrix

$$\begin{bmatrix} E & F \\ G & H \end{bmatrix} \begin{bmatrix} u_I \\ u_{II} \end{bmatrix} = \begin{bmatrix} q_I \\ -q_{II} \end{bmatrix}$$

$$E = \frac{-30\varepsilon + 10\lambda l^2 - 15\kappa l + \lambda l^6 (A-B)^2 + 5\lambda l^4 (A-B) - 10\varepsilon l^4 (A-B)^2}{30l}$$

$$F = \frac{60\varepsilon + 10\lambda l^2 + 30\kappa l + 2\lambda l^6 (A^2 - B^2) + 10\lambda l^4 A + 20\kappa l^3 A - 20\varepsilon l^4 (A^2 - B^2)}{60l} \quad (18)$$

$$G = \frac{60\varepsilon + 10\lambda l^2 - 30\kappa l + 2\lambda l^6 (A^2 - B^2) + 10\lambda l^4 A - 20\kappa l^3 A - 20\varepsilon l^4 (A^2 - B^2)}{60l}$$

$$H = \frac{-30\varepsilon + 10\lambda l^2 + 15\kappa l + \lambda l^6 (A+B)^2 + 5\lambda l^4 (A+B) - 10\varepsilon l^4 (A+B)^2}{30l}$$

Provided that the numerical values for the parameters $\varepsilon, \kappa, \lambda, l$ in the model Equation (18) are known, the element-level stiffness matrices can be evaluated. The solution of Equation (7) using bubble enriched elements is consequently achieved following the assembly of $n \times n$ global stiffness matrix and imposition of boundary conditions. The derivation of higher order least squares polynomial bubble functions can be achieved analogously by solving larger system of algebraic equations.. For example, a cubic polynomial bubble function of the form

$$\tilde{u}_j(x) = \frac{l-x}{l}u(0) + \frac{x}{l}u(l) + c_j x(l-x) + f_j x^2(l-x) \quad (19)$$

results in a linear system of two equations and two unknowns which is solved to provide

$$c_j = \frac{1}{l}\left(\left(\frac{l^7\lambda^4(u_l - 6u_0) - 40l^5\lambda^3\varepsilon(u_l - 13u_0) - 70l^5\lambda^2\kappa^2(u_l + 2u_0) - 60l^4\lambda^2\kappa\varepsilon(13u_l + 22u_0)}{l^8\lambda^4 + 52l^6\lambda^2(\kappa^2 - 2\lambda\varepsilon) + l^4(4320\lambda^2\varepsilon^2 - 1680\lambda\kappa^2\varepsilon + 420\kappa^4) + l^2\varepsilon^2(5040\kappa^2 - 60480\lambda\varepsilon) + 302400\varepsilon^4}\right)\right.$$
$$+ \left(\frac{-840l^3\lambda^2\varepsilon^2(5u_l - 16u_0) + 840l^3\lambda\varepsilon\kappa^2(-u_l + 4u_0) + 5040l^2\varepsilon^2\kappa\lambda(-u_l + 6u_0) + 2520l^2\kappa^3\varepsilon(u_l - u_0)}{l^8\lambda^4 + 52l^6\lambda^2(\kappa^2 - 2\lambda\varepsilon) + l^4(4320\lambda^2\varepsilon^2 - 1680\lambda\kappa^2\varepsilon + 420\kappa^4) + l^2\varepsilon^2(5040\kappa^2 - 60480\lambda\varepsilon) + 302400\varepsilon^4}\right)$$
$$+ \left.\left(\frac{50400l\lambda\varepsilon^3(u_l + 2u_0) + 25200lk^2\varepsilon^2(u_l - u_0) + 151200\kappa\varepsilon^3(u_l - u_0)}{l^8\lambda^4 + 52l^6\lambda^2(\kappa^2 - 2\lambda\varepsilon) + l^4(4320\lambda^2\varepsilon^2 - 1680\lambda\kappa^2\varepsilon + 420\kappa^4) + l^2\varepsilon^2(5040\kappa^2 - 60480\lambda\varepsilon) + 302400\varepsilon^4}\right)\right)$$

and

$$f_j = \frac{7}{l}\left(\left(\frac{l^6\lambda^4(-u_l + u_0) - 80l^4\lambda^3\varepsilon(-u_l + u_0) + 10l^4\lambda^2\kappa^2(-u_l + u_0) + 300l^3\lambda^2\kappa\varepsilon(u_l + u_0)}{l^8\lambda^4 + 52l^6\lambda^2(\kappa^2 - 2\lambda\varepsilon) + l^4(4320\lambda^2\varepsilon^2 - 1680\lambda\kappa^2\varepsilon + 420\kappa^4) + l^2\varepsilon^2(5040\kappa^2 - 60480\lambda\varepsilon) + 302400\varepsilon^4}\right)\right.$$
$$+ \left(\frac{1320l^2\lambda^2\varepsilon^2(-u_l + u_0) - 600l^2\lambda\varepsilon\kappa^2(-u_l + u_0) - 3600l\varepsilon^2\kappa\lambda(u_l + u_0) + 2520l^2\kappa^3\varepsilon(u_l - u_0)}{l^8\lambda^4 + 52l^6\lambda^2(\kappa^2 - 2\lambda\varepsilon) + l^4(4320\lambda^2\varepsilon^2 - 1680\lambda\kappa^2\varepsilon + 420\kappa^4) + l^2\varepsilon^2(5040\kappa^2 - 60480\lambda\varepsilon) + 302400\varepsilon^4}\right)$$
$$+ \left.\left(\frac{-7200\lambda\varepsilon^3(-u_l + u_0) + 7200\kappa^2\varepsilon^2(-u_l + u_0)}{l^8\lambda^4 + 52l^6\lambda^2(\kappa^2 - 2\lambda\varepsilon) + l^4(4320\lambda^2\varepsilon^2 - 1680\lambda\kappa^2\varepsilon + 420\kappa^4) + l^2\varepsilon^2(5040\kappa^2 - 60480\lambda\varepsilon) + 302400\varepsilon^4}\right)\right)$$

It should be noted that parameters $\varepsilon, \kappa, \lambda$ play a significant role in determining which transport phenomenon dominates the flow regime. For example, the case of $\varepsilon \ll \kappa$ represents a convection-dominated regime where it is well-known that the standard linear finite elements fail to accurately predict the near wall behaviour of solution, and hence the use of enriched numerical schemes combined with mesh refinements may be required.

## 4. Least squares bubble function for a transient model

Solution of the one-dimensional transient transport equations by using the method of partial discretization only requires spatial discretization [17]. Temporal variations in such problems are represented in a system of ordinary differential equations with given initial conditions. In this section application of the described technique to transient problems is demonstrated. Consider the following time-dependent initial-boundary value problem (also known as diffusion with lateral concentration loss problem [18])

$$\begin{cases} \dfrac{\partial u}{\partial t} + \varepsilon \dfrac{\partial^2 u}{\partial x^2} + u = 0 \\ u(x,t) = f(x), \quad t = 0 \\ u(x,t) = 0, \quad x = a, b, \quad t \geq 0 \end{cases} \tag{20}$$

A finite element discretization in the $x$ direction is applied on the equation domain $[a, b]$. Using the previous section formalisms, a trial solution is defined as

$$\tilde{u}_j(x,t) = \alpha_{1,j}\left(\dfrac{x}{l} + cx(l-x)\right) + \alpha_{2,j}\left(\dfrac{l-x}{l} + cx(l-x)\right) \tag{21}$$

on the *j-th* element, and zero elsewhere. In order to derive the unknown coefficient of the bubble function for the above expression, one may consider Equation (20) as a special case of the benchmark model (7) where $\varepsilon u'' + u = 0$. Consequently, the technique developed in the previous section can be applied in a straightforward fashion to obtain

$$c = \dfrac{-5}{2}\left(\dfrac{l^2 - 12\varepsilon}{l^4 - 20\varepsilon l^2 + 120\varepsilon^2}\right). \tag{22}$$

The weighted residual statement required for every weight function $w(x)$ is expressed as

$$0 = \int_0^l w\left(\dfrac{\partial \tilde{u}_j}{\partial t} + \varepsilon \dfrac{\partial^2 \tilde{u}_j}{\partial x^2} + \tilde{u}_j\right) dx \tag{23}$$

Application of the Green's theorem (integration by part) to (23) gives

$$\int_0^l \left(w \dfrac{\partial \tilde{u}_j}{\partial t} - \varepsilon \dfrac{\partial w}{\partial x}\dfrac{\partial \tilde{u}_j}{\partial x} + w\tilde{u}_j\right) dx = -\varepsilon \left\{w \dfrac{\partial \tilde{u}_j}{\partial x}\right\}_0^l \tag{24}$$

Corresponding to $w_0(x) = \dfrac{l-x}{l} + cx(l-x)$ and $w_1(x) = \dfrac{x}{l} + cx(l-x)$, the Galerkin weighted residual statement result in the following set of equations

$$\begin{bmatrix} \dfrac{c^2 l^6 + 5cl^4 + 10l^2}{30l} & \dfrac{c^2 l^6 + 5cl^4 + 5l^2}{30l} \\ \dfrac{c^2 l^6 + 5cl^4 + 5l^2}{30l} & \dfrac{c^2 l^6 + 5cl^4 + 10l^2}{30l} \end{bmatrix}\begin{bmatrix} \alpha_{1,j'} + \alpha_{1,j} \\ \alpha_{2,j'} + \alpha_{2,j} \end{bmatrix} + \begin{bmatrix} \dfrac{-\varepsilon(10c^2l^4 + 30)}{30l} & \dfrac{-\varepsilon(10c^2l^4 - 30)}{30l} \\ \dfrac{-\varepsilon(10c^2l^4 - 30)}{30l} & \dfrac{-\varepsilon(10c^2l^4 + 30)}{30l} \end{bmatrix}\begin{bmatrix} \alpha_{1,j} \\ \alpha_{2,j} \end{bmatrix} = \begin{bmatrix} -\varepsilon\left\{\dfrac{\partial \tilde{u}_j}{\partial x}\right\}_0^l \\ \varepsilon\left\{\dfrac{\partial \tilde{u}_j}{\partial x}\right\}_0^l \end{bmatrix}. \tag{25}$$

By setting

$$\begin{cases} \begin{bmatrix} L & M \\ M & L \end{bmatrix} \begin{bmatrix} \alpha_{1,j'} + \alpha_{1,j} \\ \alpha_{2,j'} + \alpha_{2,j} \end{bmatrix} + \begin{bmatrix} N & P \\ P & N \end{bmatrix} \begin{bmatrix} \alpha_{1,j} \\ \alpha_{2,j} \end{bmatrix} = \begin{bmatrix} q_I \\ -q_{II} \end{bmatrix} \\ L = \dfrac{c^2 l^6 + 5cl^4 + 10l^2}{30l} \\ M = \dfrac{c^2 l^6 + 5cl^4 + 5l^2}{30l} \\ N = \dfrac{-\varepsilon(10c^2 l^4 + 30)}{30l} \\ P = \dfrac{-\varepsilon(10c^2 l^4 - 30)}{30l} \end{cases} \qquad (26)$$

the assembled global stiffness equation is found as

$$\begin{bmatrix} L & M & 0 & 0 & \cdots & 0 \\ M & 2L & M & 0 & \cdots & 0 \\ 0 & M & 2L & M & \cdots & 0 \\ & & \ddots & & & \\ 0 & 0 & 0 & & M & L \end{bmatrix} \begin{bmatrix} \alpha_{1,1'} + \alpha_{1,1} \\ \alpha_{2,1'} + \alpha_{2,1} \\ \vdots \\ \alpha_{1,n'} + \alpha_{1,n} \\ \alpha_{2,n'} + \alpha_{2,n} \end{bmatrix} + \begin{bmatrix} N & P & 0 & 0 & \cdots & 0 \\ P & 2N & P & 0 & \cdots & 0 \\ 0 & P & 2L & P & \cdots & 0 \\ & & \ddots & & & \\ 0 & 0 & 0 & & P & N \end{bmatrix} \begin{bmatrix} \alpha_{1,1} \\ \alpha_{2,1} \\ \vdots \\ \alpha_{1,n} \\ \alpha_{2,n} \end{bmatrix} = \begin{bmatrix} q_I \\ 0 \\ \vdots \\ 0 \\ q_n \end{bmatrix}. \qquad (27)$$

Enforcement of the boundary conditions $u(a,t) = 0$ and $u(b,t) = 0$ and eliminating the redundant equations, corresponding to the boundary values, will result in the solution of the system and evaluation of the nodal values.

## 5. Numerical results

In this section we demonstrate the level of accuracy and reliability of the least squares bubble functions through two examples.

5.1 Numerical solution of steady-state reaction-diffusion equation

Consider the following diffusion-reaction problem (with no sources or sinks) represented as

$$\begin{cases} \dfrac{-1}{100} u'' + u = 0 \ \ in \ \Omega = [0,10] \\ u(0) = \dfrac{3}{2}, \dfrac{du}{dx}(x = 10) = 0 \end{cases} \qquad (28)$$

with the following exact solution

$$u(x) = \dfrac{3}{2} \left( \dfrac{\exp(100) \exp(-10x)}{\exp(-100) + \exp(100)} + \dfrac{\exp(-100) \exp(10x)}{\exp(-100) + \exp(100)} \right). \qquad (29)$$

This equation is typically used in modelling transport phenomena such as, for example, the decay of chemicals in a reservoir. A relatively small ratio between the diffusion and the reaction coefficient results in the trajectory of the exact solution (29) to have a very sharp gradient near the boundary wall as illustrated in Figure (1). Approximations based on Lagrangian finite elements fail to capture such an abrupt change ( the solution deteriorates rapidly as the problem becomes more reaction dominated), unless excessive mesh refinement is used in the near-boundary region.

Utilizing the described quadratic least squares bubble function we have

$$c_j = \dfrac{-25(25l^2 + 3)}{250l^4 + 50l^2 + 3}(u_l + u_0). \qquad (30)$$

The numerical result generated using bubble enriched elements for the model Equation (28), for two cases corresponding to 30 and 50 elements, is shown in Figure (1) and compared with the analytical solution. Despite using coarse mesh the bubble enriched finite element has yielded almost super-convergent solutions for this bench mark problem.

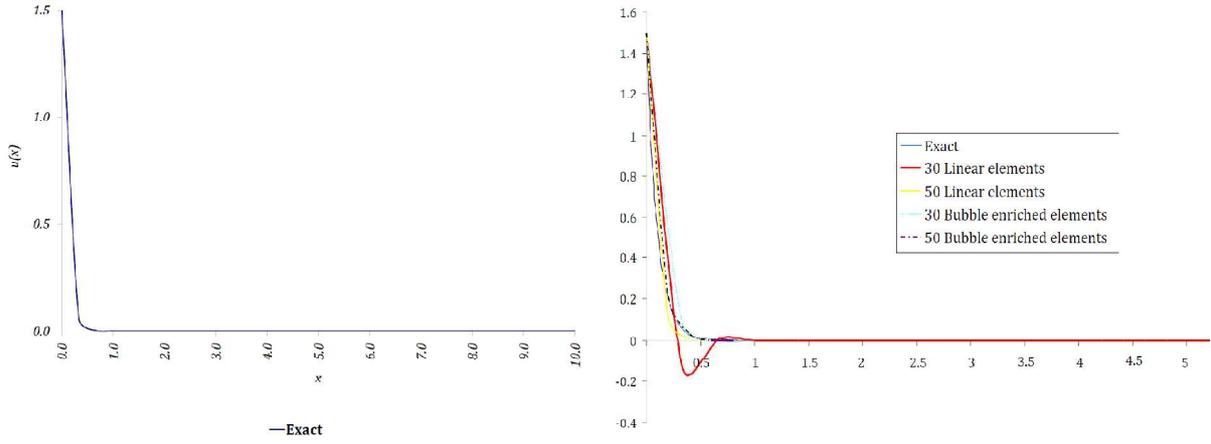

**Figure 1** Solution of the one-dimensional steady state reaction-diffusion equation; exact solution (left), comparison of the numerical solutions obtained by linear elements and bubble enriched elements against exact solution (right).

5.2 Numerical solution of transient reaction-diffusion equation

Consider the following diffusion-reaction problem over $0 \leq x \leq \pi$ expressed by

$$\begin{cases} \dfrac{\partial u}{\partial t} - \dfrac{\partial^2 u}{\partial x^2} + u = 0 \\ u(x,t) = \sin(x), \quad t = 0 \\ u(x,t) = 0 \text{ for } x = a, \pi \text{ and } 0 \leq t \end{cases} \tag{31}$$

The model governed by equation (31) is known as the heat-flow problem with lateral heat loss. The exact solution for this equation is found to be: $u(x,t) = \sin(x)\exp(-2t)$ on $x = [0, \pi]$ and $t \geq 0$.

Using a partial discretization based on a very coarse mesh consisting of only two equal size linear elements, the following solution is obtained

$$\tilde{u}_L(x,t) = \begin{cases} \dfrac{2x}{\pi} \exp(-2.216t) \text{ for } x \in \left[0, \dfrac{\pi}{2}\right] \\ \dfrac{2(\pi - x)}{\pi} \exp(-2.216t) \text{ for } x \in \left[\dfrac{\pi}{2}, \pi\right] \end{cases} \tag{32}$$

The solution for Equation (31) associated with the use of two equal size quadratic bubble-enriched elements corresponding to $\varepsilon = -1$ and $\kappa = 0$ in the benchmark model (7) is

$$\tilde{u}_B(x,t) = \begin{cases} \left(\dfrac{2x}{\pi} + 0.206x\left(\dfrac{\pi}{2} - x\right)\right) \exp(-2.031t) \text{ for } x \in \left[0, \dfrac{\pi}{2}\right] \\ \left(\dfrac{2(\pi - x)}{\pi} + 0.206(\pi - x)\left(x - \dfrac{\pi}{2}\right)\right) \exp(-2.031t) \text{ for } x \in \left[\dfrac{\pi}{2}, \pi\right] \end{cases} \tag{33}$$

Comparisons of the performance of the numerical schemes obtained from the linear elements and the quadratic bubble-enriched elements against the exact solution at $t = 0$ and $x = \dfrac{7\pi}{8}$ are given Table 1 and Table 2 and shown in Figure (2).

**Table 1** Comparison of the solution profiles for $t = 0$

| $x$ | $u(x, t = 0)$ | $\widetilde{u}_B(x, t = 0)$ | $\widetilde{u}_L(x, t = 0)$ |
|---|---|---|---|
| 0 | 0 | 0 | 0 |
| $\pi/16$ | 0.195 | 0.180 | 0.125 |
| $2\pi/16$ | 0.382 | 0.345 | 0.25 |
| $3\pi/16$ | 0.555 | 0.494 | 0.375 |
| $4\pi/16$ | 0.707 | 0.627 | 0.5 |
| $5\pi/16$ | 0.831 | 0.744 | 0.625 |
| $6\pi/16$ | 0.923 | 0.845 | 0.75 |
| $7\pi/16$ | 0.980 | 0.930 | 0.875 |
| $8\pi/16$ | 1 | 1 | 1 |
| $9\pi/16$ | 0.980 | 0.930 | 0.875 |
| $10\pi/16$ | 0.923 | 0.845 | 0.75 |
| $11\pi/16$ | 0.831 | 0.744 | 0.625 |
| $12\pi/16$ | 0.707 | 0.627 | 0.5 |
| $13\pi/16$ | 0.555 | 0.494 | 0.375 |
| $14\pi/16$ | 0.382 | 0.345 | 0.25 |
| $15\pi/16$ | 0.195 | 0.180 | 0.125 |
| $\pi$ | 0 | 0 | 0 |

**Table 2** Comparison of the solution profiles for $x = \frac{7\pi}{8}$

| $t$ | $u(x = 7\pi/8, t)$ | $\widetilde{u}_B(x = 7\pi/8, t)$ | $\widetilde{u}_L(x = 7\pi/8, t)$ |
|---|---|---|---|
| 0 | 0.382 | 0.345 | 0.25 |
| 0.1 | 0.313 | 0.281 | 0.200 |
| 0.2 | 0.256 | 0.230 | 0.160 |
| 0.3 | 0.210 | 0.187 | 0.128 |
| 0.4 | 0.171 | 0.153 | 0.103 |
| 0.5 | 0.140 | 0.125 | 0.082 |
| 0.6 | 0.115 | 0.102 | 0.066 |
| 0.7 | 0.094 | 0.083 | 0.053 |
| 0.8 | 0.077 | 0.067 | 0.042 |
| 0.9 | 0.063 | 0.055 | 0.034 |
| 1 | 0.051 | 0.045 | 0.027 |

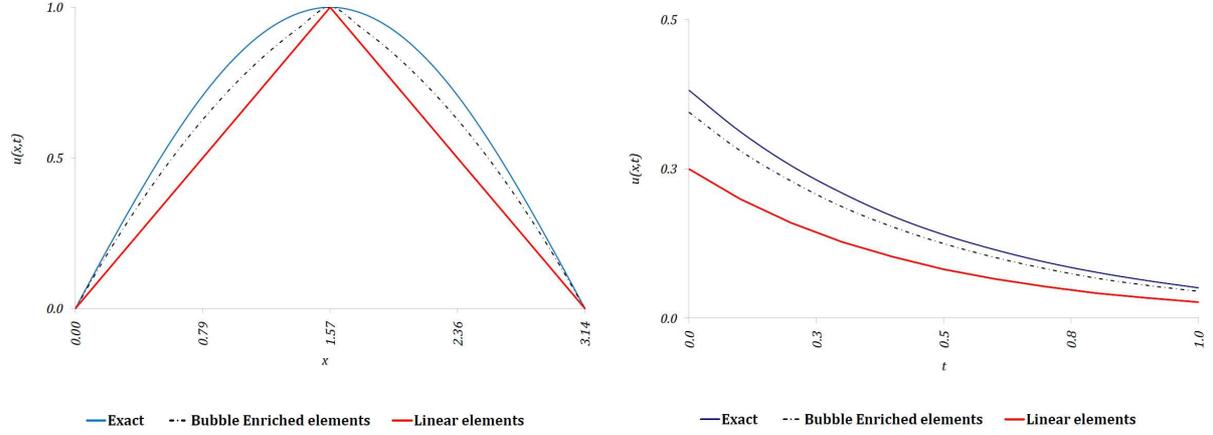

**Figure 2** Solution of the one-dimensional transient reaction-diffusion equation; comparison of the linear and bubble enriched element solutions against exact solution for the solution profile at $t = 0$ and $0 \leq x \leq \pi$ (right), and $0 \leq t \leq 1$ and $x = \frac{7\pi}{8}$ (left).

## 6. Multi-dimensional least squares bubble function enrichment

Finite element method is a particularly well suited technique for the solution of partial differential equations arising from two or three dimensional field problems defined over domains having a complex geometry [19]. The performance of the described least squares bubble-enriched scheme may consequently be tested by its application to such problems. A common procedure for the construction of multi-dimensional shape functions is the use of tensor products of one dimensional functions. However, in some extreme cases discretizations based on tensor product elements may not necessarily satisfy the original partial differential equation and its boundary conditions [20]. Consequently, techniques such as the exponential fitting or residual-free bubble function should be used. However these methods depend on finding an analytical solution at the elemental level which can prove to be impossible. In theory direct evaluation of enriched elements via the least squares method is possible but algebraic manipulations required to construct such elements can be very tedious. In this section the extension of the described method to a two dimensional example is, however, briefly presented.

Consider the following problem

$$\begin{cases} \frac{\partial u}{\partial y} = \frac{\partial^2 u}{\partial x^2} \text{ where } (x,y) \in [a,b] \times [0,\infty] \\ u(x,y) = f(x), \text{for } a \leq x \leq b \text{ and } y = 0 \\ u(x,y) = 0 \text{ for } y \geq 0 \text{ and } x = a,b \end{cases} \quad (34)$$

for which a two-dimensional polynomial bubble function is defined as $B(x,y) = \sum_{n=1}^{N}\sum_{m=1}^{M} a_n b_m x^n(l-x)y^m(h-y)$ over the rectangular master element $\Omega_{i,j} = [0,l] \times [0,h]$ in an arbitrary domain discretization. By selecting the trial approximation

$$\tilde{u}_{i,j}(x,y) = \frac{l-x}{l}\cdot\frac{h-y}{h}u(0,0) + \frac{l-x}{l}\cdot\frac{y}{h}u(0,h) + \frac{x}{l}\cdot\frac{h-y}{h}u(l,0) + \frac{x}{l}\cdot\frac{y}{h}u(l,h) + c_{i,j}xy(l-x)(h-y) \quad (35)$$

the following two-dimensional residual functional is obtained

$$J_B = \iint_{\Omega_e} R^2 \, dy \, dx \tag{36}$$

where $R = \frac{\partial^2 \tilde{u}_{i,j}}{\partial x^2} - \frac{\partial \tilde{u}_{i,j}}{\partial y}$. Minimising this functional (i.e. $\frac{\partial R}{\partial c_{i,j}} = 0$) with respect to the unknown coefficient gives

$$c = \frac{15(u(0,0) - u(0,h) + u(l,0) - u(l,h))}{h(l^4 + 12h^2)} \tag{37}$$

Least squares bubble functions can similarly be evaluated for other equation types and higher order polynomials.

## 7. Discussion and conclusion

In this paper, a least squares method for the derivation of polynomial bubble functions is proposed and is used as a uniform approximation for residual-free bubble functions to enrich standard linear finite elements in the Galerkin weighted-residual statements of multiscale problems. Quadratic and cubic least squares bubble functions are derived for a generic one-dimensional steady-state transport model and the generalization of the method, using partial discretization, to transient transport models has been demonstrated. The practicality and satisfactory performance of this method is tested by numerical examples and comparing the obtained results against the exact solution. It is shown that the best least squares bubble functions are derived and implemented following relatively straightforward calculations and in general, they have advantages over the standard schemes. These elements are typically required to capture the abrupt changes in the solution of multiscale problems. In contrast to residual free approach for the enrichment of finite elements the described method can be fully automated for implementation in finite element codes. Derivation of residual free bubble functions requires analytical solution of the model differential equations at an elemental. Therefore, in practice, unrealistic approximations may be needed to derive them. As the presented least squares method does not have such a drawback further extension of this method may yield powerful schemes for the solution of a wide range of multiscale transport problems.

**Acknowledgement**
The first author wishes to thank the Department of Chemical Engineering at Loughborough University for the study grant that made this work possible.

**References**

[1] J. Petera, V. Nassehi, J. F. T. Pittman; Petrov-Galerkin methods on isoparametric bilinear and biquadratic elements tested for a scalar convection-diffusion problem, International Journal for Numerical and Heat Fluid Flow, 3 (1993) 205-221.

[2] M. Parvazinia, V. Nassehi, R. J. Wakeman, M. H. R. Ghoreishy; Finite element modelling of flow through a porous medium between two parallel plates using the Brinkman equation, Transport in Porous Media, 63 (2006) 71-90.

[3] M. Parvazinia, V. Nassehi; Using bubble functions in the multi-scale finite element modelling of the convection-diffusion-reaction equation, International Communications in Heat and Mass Transfer, 37 (2010) 125-130.

[4] T. J. R. Hughes; Multi-scale phenomena; Green's function, the Dirichlet-to-Neumann formulation, subgrid scale models, bubbles and the origins of stabilized methods, Comput. Methods Appl. Mech. Engrg., 127 (1995) 387-401.

[5] T. J. R. Hughes, G. R. Feijo'o, L. Mazzei, J. B. Quincy; The variational Multi-scale method- a paradigm for computational mechanics; Comput. Methods Appl. Mech. Engrg., 166 (1998) 3-24.

[6] T. J. R. Hughes, J. Stewart; A space time formulation for multiscale phenomena, Journal of Computational and Applied Mathematics, 74 (1-2) (1996) 217-229.

[7] D.N. Arnold, F. Brezzi, M. Fortin; A stable finite element for the Stokes equations, Calcolo, 21, 4 (1984) 337-344.

[8] F. Brezzi, L. P. Franca, T. J. R. Hughes, A. Russo; $b=\int g$, Comput. Methods Appl. Mech. Engrg., 145 (1997) 339-392.

[9] F. Brezzi, M. Bristeau, L. P. Franca, M. Mallet; A relationship between stabilized finite element methods and the Galerkin method with bubble functions, Comput. Methods Appl. Mech. Engrg., 96 (1) (1992) 117-129.

[10] C. Baiocchi, F. Brezzi, L. P. Franca; Virtual bubbles and Galerkin-least-squares type methods, Comput. Methods Appl. Mech. Engrg., 105 (1) (1993) 125-141.

[11] J. Donea, A. Huerta; Finite Element Method for Flow Problems, 2003, John Wiley & Sons Ltd.

[12] M. Parvazinia, V. Nassehi, R.J. Wakeman; Multi-scale finite element modelling using bubble function method for a convection-diffusion problem, Chemical Engineering Science, 61 (2006) 2742-2751.

[13] M. Parvazinia, V. Nassehi; Multi-scale finite element modelling of diffusion-reaction equation using bubble functions with bilinear and triangular elements, Comput. Methods Appl. Mech. Engrg., 196 (2007) 1095-1107.

[14] L. P. Franca, A. Russo; Unlocking with residual-free bubbles, Comput. Methods Appl. Mech. Engrg. 142 (1997) 361-364.

[15] L. P. Franca, A. L. Madureira, F. Valentin; Towards multiscale functions: enriching finite element spaces with local but not bubble-like functions, Comput. Methods Appl. Mech. Engrg. 142 (1997) 361-364.

[16] L. Collatz; The Numerical Treatment of Differential Equations, 1960, Springer, Berlin.

[17] O. C. Zienkiewicz, K. Morgan; Finite Element and Approximation, 1983, Wiley, New York.

[18] S. J. Farlow; Partial Differential Equations for Scientists and Engineers, 1993, Dover, New York.

[19] V. Nassehi; Practical Aspects of Finite Element Modelling of Polymer Processing, 2002, Wiley, Chichester.

[20] L. P. Franca, A. L. Madureira, L. Tobiska, and F. Valentin; Convergence Analysis of a Multiscale Finite Element Method for singularly Perturbed Problems, SIAM Journal of Multiscale Modelling and Simulation 4, No. 3, (2005) 839-866.